\documentclass[prl,aps,twocolumn,epsfig]{revtex4}
\usepackage{graphicx}
\usepackage{dcolumn}
\usepackage{bm}
\makeatletter

\newcommand{\Rmnum}[1]{\expandafter\@slowromancap\romannumeral #1@}
\makeatother
\begin{document}
\title {Spontaneous breaking of time reversal symmetry in strongly interacting two dimensional electron layers in silicon and germanium}
\author{S. Shamim,$^1$ S. Mahapatra,$^2$ G. Scappucci,$^2$ W.M.Klesse,$^2$ M.Y.Simmons,$^2$ and A. Ghosh,$^1$}
\vspace{1.5cm}
\address{$^1$ Department of Physics, Indian Institute of Science, Bangalore 560 012, India}
\address{$^2$ Centre for Quantum Computation and Communication Technology, University of New South Wales, Sydney NSW 2052, Australia}

\begin{abstract}
We report experimental evidence of a remarkable spontaneous time reversal symmetry breaking in two dimensional electron systems formed by atomically confined doping of phosphorus (P) atoms inside bulk crystalline silicon (Si) and germanium (Ge). Weak localization corrections to the conductivity and the universal conductance fluctuations were both found to decrease rapidly with decreasing doping in the Si:P and Ge:P $\delta-$layers, suggesting an effect driven by Coulomb interactions. In-plane magnetotransport measurements indicate the presence of intrinsic local spin fluctuations at low doping, providing a microscopic mechanism for spontaneous lifting of the time reversal symmetry. Our experiments suggest the emergence of a new many-body quantum state when two dimensional electrons are confined to narrow half-filled impurity bands.
\end{abstract}

\maketitle

Invariance to time reversal is among the most fundamental and robust symmetries of nonmagnetic quantum systems. Its violation often leads to new and exotic phenomena, particularly in two dimensions (2D), such as the quantized Hall conductance in semiconductor heterostructures~\cite{Klitzing_PRL1980}, the quantum anomalous Hall effect in topological insulators~\cite{Chang_TI_Science2013} or the predicted chiral superconductivity in graphene~\cite{Levitov_NatPhys2012}. The breaking of time reversal invariance is experimentally achieved either by an external magnetic field or intentional magnetic doping. Here we show that strong Coulomb interactions can also lift the time reversal symmetry in nonmagnetic 2D systems at zero magnetic field.

While bulk P-doped Si and Ge have been extensively studied in the context of electron localization in three dimensions~\cite{Rosenbaum_PRB1983,Sarachik_PRB1992,Loh_PRL1989,Sachdev_Paalanen_PRL1988,Sachdev_Localmoment_PRB1989,
Sachdev_Bhatt_LocalMoment_PRL1989}, confining the dopants to one or few atomic planes ($\delta-$layers) of the host semiconductor has recently led to a new  class of 2D electron system~\cite{Goh2006,Giordano_Nanolett2012,BentScience,Fuechsle2012}. Electron transport in these atomically confined 2D layers occurs within a 2D impurity band where the effective Coulomb interaction is parameterized in terms of $U/\gamma$, with $U$ being the Coulomb energy required to add an additional electron to a dopant site, and $\gamma$, the hopping integral between adjacent dopants. Since each dopant P atom contributes one valence electron, the impurity band is intrinsically 'half filled' (schematic in Fig.~1a), which reinforces the interaction effects due to the in-built electron-hole symmetry, and forms an ideal platform to explore the rich phenomenology of the 2D Mott-Hubbard model, ranging from Mott metal-insulator transition (MIT) to novel spin excitations and magnetic ordering~\cite{Vollhardt_PRL2005,NandiniTrivedi_PRL1999,Kohno_PRL2012,BhatPRB2007}.

In this Letter we show evidence of spontaneously broken time reversal symmetry in 2D Si:P and Ge:P $\delta$-layers as the on-site effective Coulomb interaction is increased by decreasing the doping density of P atoms. Quantum transport and noise experiments indicate a strong suppression of quantum interference effects at low doping densities. We could attribute this to a spontaneous breaking of time reversal symmetry which manifest in an unambiguous suppression of universal conductance fluctuations (UCF) at zero magnetic field.

The preparation of the P $\delta$-layers in Si and Ge have been detailed in earlier publications~\cite{Goh2006,Johnson_PhDThesis,Giordano_Nanolett2012}, and parameters relevant to the present work is supplied in the Supplementary Information (SI). The Drude conductivity ($\sigma_D$) of the $\delta$-layers decreases with decreasing doping as $\sigma_D \propto n^{3/2}$ (Fig.~1b), where $n$ is the electron density measured from Hall effect, implying significant scattering from charged dopants~\cite{SDSarma_SiP_PRB2013}. We find $\sigma_D \gg e^2/h$ in all devices, ensuring a nominally weakly localized regime. All electrical transport measurements were carried out in a dilution refrigerator with an electron temperature of $0.15$~K using low frequency ac lock-in technique.  The electron transport was strictly diffusive with $k_BT\tau_0/\hbar \ll 10^{-2}$, because of short momentum relaxation times  $\tau_0 \sim 10 - 100$~fs, and displays negative logarithmic correction to conductivity in the quantum coherent regime (Fig.~1c)~\cite{Giordano_Nanolett2012}.

\begin{figure}[t!]
\includegraphics[width=1\linewidth]{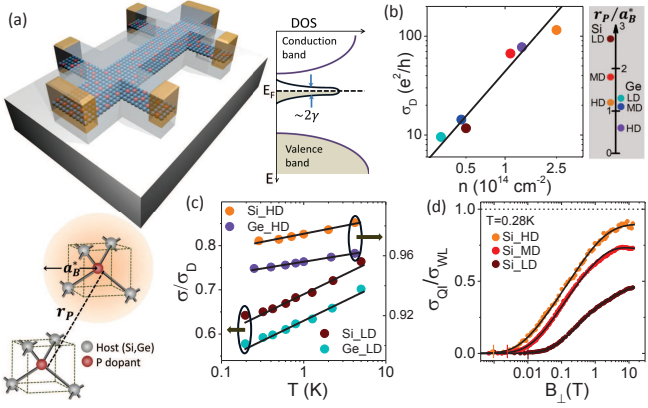}
\caption{(Color online) (a) Schematic showing the 2D device architecture, incorporation of P atoms in Si/Ge tetrahedra ($a_B^*$ is the effective density-of-states of Bohr radius and $r_P$is the dopant separation) and the band diagram . The Fermi energy, $E_F$, lies near the center of the impurity band whose width is determined by the hopping integral, $\gamma$. (b) The Drude conductivity $\sigma_D$, as a function of $n$ for SiP and GeP devices. The range of the effective dopant separation, $r_P/a_B^*$, and the device nomenclature are shown in the shaded panel on the right, where HD, MD and LD correspond to high density, medium density and low density respectively. The corresponding densities are $2.5, 1.1$ and $0.5\times10^{14}$~cm$^{-2}$ respectively for Si and $1.35, 0.46$ and $0.32\times10^{14}$~cm$^{-2}$ respectively for Ge. (c) The temperature dependence of conductivity, $\sigma$ (scaled by the Drude conductivity, $\sigma_D$) for heavily and lightly doped $\delta$-layers in Si and Ge. (d) The quantum correction to conductivity. $\sigma_{QI}$ (obatined from measured magnetoconductivity after eliminating the classical contribution) as a function of perpendicular magnetic field, $B_\perp$, at $0.28$~K for Si\_HD, Si\_MD and Si\_LD. The phase breaking field, $B_\phi$, is shown by vertical lines. The solid black lines are fits using Eq.~1 in the main text.}
\end{figure}

The key advantage of using both Si and Ge as host semiconductors is the factor of three difference in the Bohr radius, $a_B^*$, which allows us to achieve a wide range of average effective dopant separation ($r_P/a_B^*$) within the similar range of doping density ($r_P \approx 2/\sqrt{\pi n}$). As shown in the scale bar of Fig.~1b, $r_P/a_B^*$ has an overall range from $\approx 0.6$ to 3. This corresponds to a range of $\gamma \sim 10-20$~meV and $\sim 20-50$~meV for the Ge:P and Si:P devices respectively, assuming hydrogenic orbitals~\cite{Efros_Shklovskii_Book}. Since $U \sim 200$~meV and $\sim 50$~meV for single P donor in Si and Ge, respectively, the effective on-cite Coulomb interaction $U/\gamma$ can be $\gg 1$, particularly in lightly doped Si devices.

\begin{figure}[t]
\begin{center}
\includegraphics[width=1\linewidth]{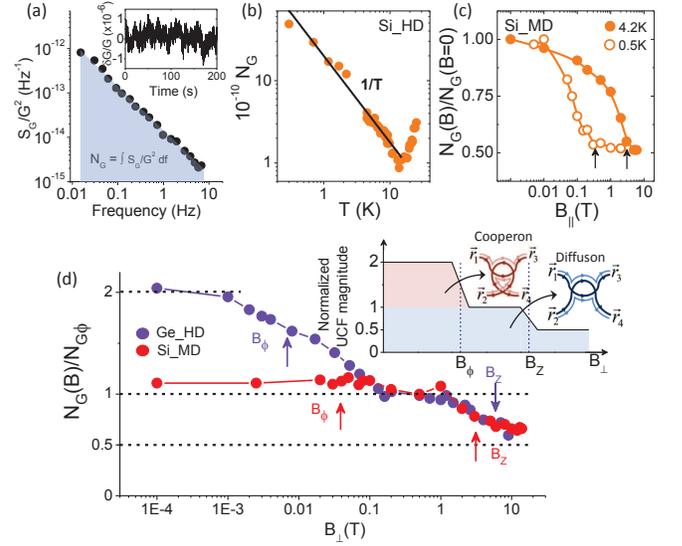}
\caption{(Color online) (a) Typical power spectral density of conductance fluctuations, $S_G$. The shaded region represents the normalized variance given by $N_G = \int S_G/G^2 df = \langle\delta{G^2}\rangle/\langle{G^2}\rangle$.  Inset shows the normalized conductance fluctuations ($\delta{G}/G$) in real time. (b) $N_G$ as a function of temperature $T$ for Si\_HD. The dotted line shows that noise $\sim 1/T$ in the low $T$ regime. (c) $N_G(B)/N_G(B=0)$ for Si\_MD as a function of parallel magnetic field, $B_\|$, at $4.2$~K and $0.5$~K. The vertical arrows denote $B_Z$. (d) $N_G(B_\perp)/N_{G\phi}$ for Ge\_HD and Si\_MD as a function of perpendicular magnetic field, $B_\bot$, at $4.2$~K where $N_{G\phi}=N_G$($B_\perp \sim 20B_\phi$). Inset is the schematic showing the reduction in UCF magnitude by factors of two at two characteristic field scales, $B_\phi$ and $B_Z$ (shown by vertical arrows). $B_\phi$ and $B_Z$ are the phase breaking field obtained from low field magnetoconductivity fits and the Zeeman field respectively. }
\end{center}
\vspace{-0.3cm}
\end{figure}

In Fig.~1d, we show the transverse magnetic field ($B_\perp$) dependence of the quantum correction to conductivity, $\sigma_{QI}(B_\perp)=\sigma(B_\perp)-\sigma(0)-\sigma_{cl}$, where $\sigma_{cl} = -(\sigma_D^3/n^2e^2)B_\perp^2$, is the classical correction to the Drude conductivity. Due to diffusive nature of our devices the quantum correction from the electron-electron interaction is only perturbative ($\sim (\omega_c\tau_0)^2$ $\lesssim 10^{-4}$, where $\omega_c$ is the cyclotron frequency)~\cite{Interaction_MR_PRL} and $\sigma_{QI}(B_\perp)$ represents the contribution primarily from the quantum interference effect. $\sigma_{QI}$ for three 2D Si:P $\delta$-layers at $0.28$~K is shown in Fig.~1d. For comparison, $\sigma_{QI}$ is scaled by $\sigma_{WL}$, where  $\sigma_{WL} = (e^2/\pi h)\ln{(\tau_\phi/\tau_0)}$ is the universal weak localization correction to conductivity for a diffusive 2D conductor with free electrons. For each device, both $\sigma_{WL}$ and the phase breaking field $B_\phi = \hbar/4eD\tau_\phi$ (shown by vertical lines in Fig.~1d) were experimentally estimated from the low-$B_\bot$ magnetoconductivity data (see SI, section S3), where $\tau_\phi$ and $D$ are the phase coherence time and electron diffusivity, respectively. Since the magnitude of $\sigma_{QI}$ at $B \gg B_\phi$ represents the net correction to conductivity due to quantum interference, it is evident from Fig.~1d that the contribution of weak localization effect on transport decreases with decreasing doping density (see SI, section S1). It is important to note that a major shift in the dominant dephasing mechanism in the lightly doped samples is ruled out because we find $\tau_\phi$ to be similar in magnitude in all three devices, and $\propto T$ down to $T = 0.2$~K (Fig.~S2 in SI). This confirms the predominance of the electron-electron scattering mediated dephasing which has been reported earlier in such $\delta$-layers~\cite{Giordano_Nanolett2012}.

The reduced quantum correction cannot be due to finite experimental range ($\approx 0-14$~T) of $B_\perp$, which exceeds both $B_\phi$ and $B_0$ ($=\hbar/4eD\tau_0$, the upper cutoff field due to momentum relaxation) by factors of $1000$ and $2$ respectively even for the least doped devices at $0.28$~K (Table I in SI). Spin-orbit interaction is also known to be small for P-doped (bulk) Si and Ge~\cite{AGPRL2000,Critical_exponent_SiP_PRB}, and independent of  the density of the dopants. Any long range magnetic order is also unlikely because the Hall resistance was found to vary linearly with $B_\bot$ at all $T$ (see SI, section S7) in all our devices~\cite{BhatPRB2007}.

The suppression of quantum correction to conductivity has been observed in low density electron gases in Si MOSFETs near the apparent MIT~\cite{Kravcehnko_PRL2003} although its microscopic origin remains unclear with both temperature dependant screening of disorder and interaction driven spin fluctuations suggested as competing mechanisms. However, the formation of local magnetic moments in the presence of strong Coulomb interactions, is known to occur in three dimensional P-doped Si close to the MIT~\cite{Sachdev_Paalanen_PRL1988,Sachdev_Localmoment_PRB1989,
Sachdev_Bhatt_LocalMoment_PRL1989}. These moments serve to remove the time reversal symmetry, suppressing the coherent back-scattering of electrons. In 2D, the possibility of localized spin excitations at the Mott transition has been suggested theoretically~\cite{Kohno_PRL2012,Imada_RMP1998}, but without any experimental evidence so far.

\begin{figure}[t]
\includegraphics[width=1\linewidth]{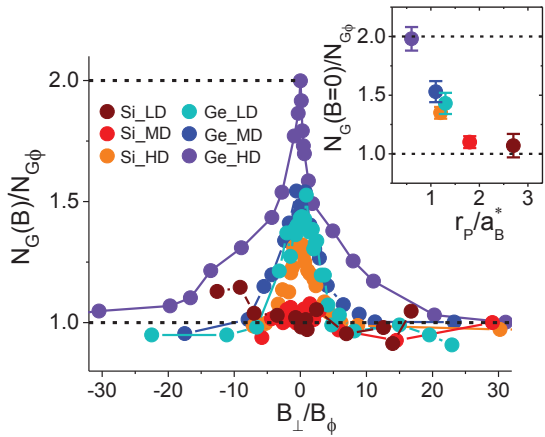}
\caption{(Color online) (a) $N_G(B_\bot)/N_{G\phi}$ as a function of $B_\bot$ (scaled by the phase breaking field, $B_\phi$) for all devices at $4.2$~K, where $N_{G\phi}=N_G$($B_\perp \sim 20B_\phi$). The inset shows $N_G(B_\perp=0)/N_{G\phi}$ as a function of $r_P/a_B^*$.}
\end{figure}

To probe whether the observed suppression of localization correction indeed manifests a breaking of the time reversal symmetry, we have measured the UCF as a function of  $T$ and $B_\perp$ from slow time-dependent fluctuations in the conductance ($G$) of the $\delta$-layers which represents the ensemble fluctuations via the ergodic hypothesis~\cite{Stone_PRB,AGPRL2000,BirgePRB1990,BirgePRB1993,FengPRL1986}. The time dependant conductance fluctuations (inset of Fig.~2a) are analyzed to obtain the power spectral density, $S_G$, which on integration over the experimental bandwidth gives the normalized variance, $N_G=\int{S_G/G^2}df=\langle\delta{G}^2\rangle/\langle{G}\rangle^2$ as shown in Fig.~2a (see Ref~\cite{SaquibPRB2011} and SI, section S3 for details). Fig.~2b shows $N_G$ as a function of $T$ for Si\_HD. For $T \lesssim 15$~K, $N_G$ increases with decreasing $T$, which is a hallmark of UCF. In this regime, one expects $N_G \propto L_\phi^4n_T \propto 1/T$, where $L_\phi (\propto T^{-0.5})$ and $n_T (\propto T)$ are the phase coherence length and density of active two level fluctuators~\cite{BirgePRB1990} (Fig.~2b). The absolute magnitude of $N_G$ in all devices correspond to the change in conductance by $\sim O[e^2/h]$ due to a single fluctuator within a phase coherent box (see SI, section S5), establishing the observed noise to be indeed from mesoscopic fluctuations.

As a function of $B_\perp$, the magnitude of UCF is expected to decrease by an exact factor of two at two field scales, first at $B_\perp \sim B_\phi$ when the time reversal symmetry, and hence the Cooperon (self-intersecting diffusion trajectories) contribution, is removed~\cite{Altshuler_Spivak_1985,Stone_PRB,Beenakker_Houten_Review} and second at $B_\perp \sim B_Z = k_BT/g\mu_B$ due to removal of spin degeneracy~\cite{Stone_PRB,BirgePRB1996_Zeeman,Beenakker_Houten_Review}, where $g$ and $\mu_B$ are the $g$-factor and $\mu_B$ respectively. The inset of Fig.~2d shows schematically the two reductions in UCF magnitude as a function of $B_\perp$.  Fig.~2d shows that the UCF magnitude in heavily doped Ge\_HD (violet symbols) consists of both factors of two reduction at $B_\bot \approx B_\phi$ and $B_\bot \approx B_Z$, corresponding to the removal of time reversal symmetry and spin degeneracy, respectively, whereas the lightly doped devices, such as Si\_MD, shows almost no variation in the UCF magnitude on the scale of $B_\phi$ but decreases by a factor of two at $B_\bot \approx B_Z$. To confirm this scenario, we have also recorded the variation of $N_G$ in Si\_MD as a function of parallel magnetic field, $B_\|$, which couples only to spin degree of freedom (Fig.~2c). The factor of two reduction at $B_\| \sim B_Z$ (shown by vertical arrows in Fig.~2c) for $T=0.5$~K and $4.2$~K establishes that the $1/f$ noise in our devices indeed arises from the UCF mechanism.

Since the reduction in UCF at $B_\bot \sim B_\phi$ is associated only to removal of the fundamental time reversal symmetry of the underlying Hamiltonian~\cite{Altshuler_Spivak_1985}, its absence in the lightly doped $\delta-$layers is unique, and has not been previously observed in interacting 2D systems in semiconductors~\cite{MYS_NatPhys2008,Kravchenko_RMP2001}. To elaborate, we have compiled the $B_\bot$-dependence of $N_G$ normalized by $N_{G\phi}$, where $N_{G\phi}$ is the value of $N_G$ at $B_\perp \gg B_\phi$ but $< B_Z$ , for all devices in Fig.~3. $N_{G\phi}$ was chosen at $B_\perp \sim 20B_\phi$ which was $< B_Z$ for all the devices at all temperatures. The peak in $N_G$ around $B_\bot = 0$ is progressively suppressed with decreasing doping density, and eventually for $r_P/a_B^* \gtrsim 1.5$, the Cooperon contribution to UCF noise at low $B_\perp$ becomes immeasurably small, implying a spontaneous breaking of time reversal symmetry even at $B_\perp = 0$ (Inset of Fig.~3).

\begin{figure}[t]
\includegraphics[width=1\linewidth]{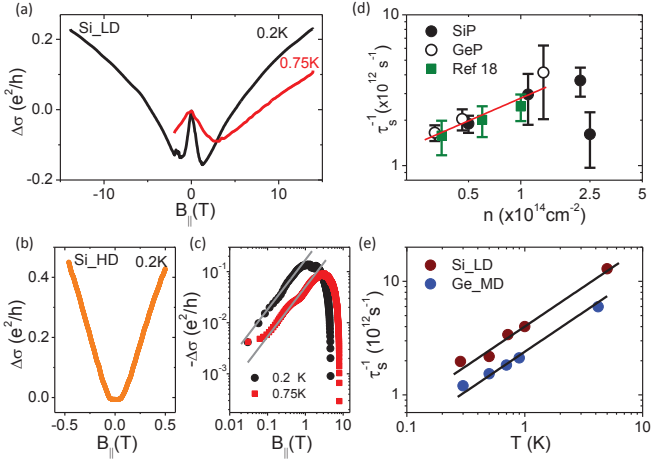}
\caption{(Color online) (a) The magnetoconductivity, $\Delta\sigma$ in presence of magnetic field, $B_\parallel$, applied parallel to the plane of the $\delta$-layer for Si\_LD at $0.2$~K and $0.75$~K. (b) $\Delta\sigma$ in $B_\parallel$ for Si\_HD at $0.2$~K. (c) $\Delta\sigma$ as a function of $B_\parallel$ for Si\_LD at $0.2$~K and $0.75$~K in log-log scale. The solid lines show that $\Delta\sigma\propto B_\parallel$ in the region of negative magnetoconductivity. (d) The spin scattering rate, $\tau_s^{-1}$, as a function of carrier density, $n$, for all devices at $0.28$~K. The solid line shows that $\tau_s^{-1} \propto n^{0.5}$. (e) $\tau_s^{-1}$ as a function of $T$ for Si\_LD and Ge\_MD. The solid lines show that $\tau_s^{-1} \propto T^{0.7}$ for both the devices. }
\end{figure}

To explore the origin of lifting of the time reversal symmetry in the $\delta$-layers, we subjected the devices to {\it in-plane} magnetic field, $B_\|$, that resulted in a nonmonotonic magnetoconductivity in the lightly doped $\delta$-layers. The logarithmic increase in the magnetoconductivity at large $B_\|$, as shown in Fig.~4a, was observed in all devices irrespective of doping level, and known to represent suppression of weak localization due to the finite width of the $\delta$-layers~\cite{InplaneMR_Meyer_arxiv}. However, the negative magnetoconductivity around $B_\| = 0$ often indicates the presence of local moments, because localization strengthens as phase coherence increases with the freezing of spin-flip scattering~\cite{MR_CeRhIn_PRB2002,InplaneMR_Meyer_arxiv}. In such a case, the activated spin-flip processes across the Zeeman gap, leads to magnetoconductivity decreasing linearly with $B_\|$ as $\Delta\sigma(B_\|) = -\eta B_\|/T$, where $\eta\sim e^2g_{imp}\mu_B/hk_B$, and $g_{imp}$ is the $g$-factor of the magnetic impurity~\cite{InplaneMR_Meyer_arxiv}. As shown in Fig.~4c, we indeed find the $\Delta\sigma(B_\|,T) \propto B_\|/T$ in Si\_LD. The negative magnetoconductivity in $B_\|$ is entirely absent in the heavily doped devices (Fig.~4b). This establishes that the spin fluctuations are entirely due to strong Coulomb interactions, and hence observable only in the lightly doped $\delta$-layers. Importantly, the experimental value of $\eta$ was found to be a factor of $\sim 50$ smaller than that expected theoretically (assuming $g_{imp}=2$), suggesting that the impact of local moments on the dephasing process is anomalously small.

The compelling analogy with the bulk P-doped Si close to MIT provides a ``two-fluid'' framework to address transport in our $\delta$-layers. This consists of itinerant electrons in disordered Hamiltonian and local magnetic moments~\cite{Sachdev_Paalanen_PRL1988,Sachdev_Localmoment_PRB1989,
Sachdev_Bhatt_LocalMoment_PRL1989}. The interaction between the local moments and itinerant electrons suppresses localization, although the spin-scattering process is quasi-elastic (energy exchange $\ll k_BT$), causing only minor modification to the dephasing mechanism (as confirmed by the linear $T$ dependence of $\tau_\phi^{-1}$ in Fig.~S2 of SI and small $\Delta\sigma(B_\|)$). In addition, the two-fluid model allows a phenomenological generalized Hikami-Larkin-Nagaoka expression for the total quantum interference correction that includes the quasi-elastic spin scattering rate ($\tau_s^{-1}$) as,

\begin{eqnarray}
\label{deltasigma_total}
\Delta\sigma(B_\bot,T)=\frac{\alpha e^2}{\pi h}\left[F\left(\frac{B_\bot}{B_\phi}\right) - F\left(\frac{B_\bot}{B_0}\right)\right] - \frac{\beta e^2}{\pi h} F\left(\frac{B_\bot}{B_s}\right)
\end{eqnarray}

\noindent where $\alpha$ and $\beta$ are positive constants close to unity, and $F(x) = \ln(x)+\psi(0.5+1/x)$, with $\psi(x)$ being the digamma function. As shown by the solid lines in Fig.~1d, Eq.~\ref{deltasigma_total} describes the magnetoconductivity very well over the entire range of $B_\bot$. The fit parameter $B_s = \hbar/4eD\tau_s$, provides an estimate of the spin scattering time $\tau_s$. We note the following: (i) As evident in Fig.~4d, $\tau_s^{-1}$ is more than ten times larger than experimentally measured $\tau_\phi^{-1}$ (see SI), confirming that the spin-scattering is mostly elastic. (ii) Second, $\tau_s^{-1}$ varies nonmonotonically with $n$. The filled squares represent $\tau_s^{-1}$ analyzed from data of Ref~\cite{Johnson_PhDThesis}. At low $n$, $\tau_s^{-1} \sim n^{0.5}$ irrespective of host material, disorder or carrier mobility, indicating that the number of local spins are only related with the number of P dopant sites. However $\tau_s^{-1}$ drops abruptly around $n \sim 1.5\times10^{14}$~cm$^{-2}$, suggesting a quenching of the spins and commencement of free-electron weakly localized quantum transport. The T-dependence of $\tau_s^{-1}$ (Fig.~4e), in accordance with the two-fluid model, shows a power law variation as $\tau_s^{-1} \propto T^p$, with $p \approx 0.7$. This sets the exponent for susceptibilty and specific heat divergence in the $\delta$-layers to be $\approx 0.3$, which is about half of that observed in the bulk Si:P close to MIT~\cite{Sachdev_Paalanen_PRL1988,Lohneysen_SpecificHeat_PRL1989}.

Finally, to estimate the fraction of P-dopants that host a local moment, we compare the estimated $\tau_s^{-1}$ in lightly doped Si\_LD ($n = 5\times10^{13}$~cm$^{-2}$) with (1) the total momentum relaxation rate $\tau_0^{-1} \approx 10^{14}$~s$^{-1}$ from the experimental Drude conductivity, although this involves scattering from neutral defects as well, and (2) calculated momentum relaxation rate ($\approx 2\times10^{13}$~s$^{-1}$) expected purely from the P-dopants (charged impurities) (see calculation details in Ref~\cite{SDSarma_SiP_PRB2013} and SI, section S6). This gives a bound between $2\% - 10\%$ of the P-dopants to host local moments which is consistent with the fraction expected for half-filled impurity bands in bulk Si:P~\cite{Sachdev_Bhatt_LocalMoment_PRL1989}. Importantly, while the weak localization correction is reduced only partially (~30\% in Si\_LD), the UCF noise due to the Cooperons is completely suppressed for the weakly doped devices. It is possible that because the UCF noise involves interference between two Feynman propagators, it is more likely to be affected by the localized spins than the WL correction which is determined by a single self intersecting propagator. Note that we have not discussed spatial inhomogeneity or clustering in the distribution of dopants which can lead to coexistence of localized and delocalized phases~\cite{Vollhardt_PRL2005}, impact of multiple valleys~\cite{Punnoose_Science2005,Medini_NatPhys2006}, or the inter-site Coulomb interaction~\cite{Shepelyansky_PRL1994,MYS_NatPhys2008,Kravchenko_RMP2001} which are unlikely to affect the time reversal symmetry.

In summary, magnetoconductivity and noise measurements reveal an unexpected spontaneous breaking of time reversal symmetry in 2D electron systems hosted in atomically confined Si:P and Ge:P crystals. The universal conductance fluctuations and in-plane magnetoconductivity suggest that local spin fluctuations in the presence of strong Coulomb interaction play an important role in the lifting the time reversal symmetry. Whether this indeed leads to a true interaction-induced metallic ground state in two dimensions needs further experimental and theoretical exploration.

\section{Acknowledgement}

We acknowledge Sankar Das Sarma, Ravin N.Bhatt, Vijay Shenoy, Sanjoy Sarker and Jainendra Jain for discussions. We thank Department of Science and Technology (DST), Government of India and Australian-Indian Strategic Research Fund (AISRF) for funding the project. The research was undertaken in collaboration with the Australian Research Council, Centre of excellence for Quantum Computation and Communication Technology (Project number CE110001027) and the US Army Research Office under contract number W911NF-08-1-0527. SS thanks CSIR for financial support. GS acknowledges support from UNSW under the GOLDSTAR Award 2012 scheme. MYS acknowledges a Federation Fellowship.


\begin{thebibliography}{41}
\expandafter\ifx\csname natexlab\endcsname\relax\def\natexlab#1{#1}\fi
\expandafter\ifx\csname bibnamefont\endcsname\relax
  \def\bibnamefont#1{#1}\fi
\expandafter\ifx\csname bibfnamefont\endcsname\relax
  \def\bibfnamefont#1{#1}\fi
\expandafter\ifx\csname citenamefont\endcsname\relax
  \def\citenamefont#1{#1}\fi
\expandafter\ifx\csname url\endcsname\relax
  \def\url#1{\texttt{#1}}\fi
\expandafter\ifx\csname urlprefix\endcsname\relax\def\urlprefix{URL }\fi
\providecommand{\bibinfo}[2]{#2}
\providecommand{\eprint}[2][]{\url{#2}}

\bibitem[{\citenamefont{Klitzing et~al.}(1980)\citenamefont{Klitzing, Dorda,
  and Pepper}}]{Klitzing_PRL1980}
\bibinfo{author}{\bibfnamefont{K.~v.} \bibnamefont{Klitzing}},
  \bibinfo{author}{\bibfnamefont{G.}~\bibnamefont{Dorda}}, \bibnamefont{and}
  \bibinfo{author}{\bibfnamefont{M.}~\bibnamefont{Pepper}},
  \bibinfo{journal}{Phys. Rev. Lett.} \textbf{\bibinfo{volume}{45}},
  \bibinfo{pages}{494} (\bibinfo{year}{1980}).

\bibitem[{\citenamefont{Chang et~al.}(2013)\citenamefont{Chang, Zhang, Feng,
  Shen, Zhang, Guo, Li, Ou, Wei, Wang et~al.}}]{Chang_TI_Science2013}
\bibinfo{author}{\bibfnamefont{C.-Z.} \bibnamefont{Chang}},
  \bibinfo{author}{\bibfnamefont{J.}~\bibnamefont{Zhang}},
  \bibinfo{author}{\bibfnamefont{X.}~\bibnamefont{Feng}},
  \bibinfo{author}{\bibfnamefont{J.}~\bibnamefont{Shen}},
  \bibinfo{author}{\bibfnamefont{Z.}~\bibnamefont{Zhang}},
  \bibinfo{author}{\bibfnamefont{M.}~\bibnamefont{Guo}},
  \bibinfo{author}{\bibfnamefont{K.}~\bibnamefont{Li}},
  \bibinfo{author}{\bibfnamefont{Y.}~\bibnamefont{Ou}},
  \bibinfo{author}{\bibfnamefont{P.}~\bibnamefont{Wei}},
  \bibinfo{author}{\bibfnamefont{L.-L.} \bibnamefont{Wang}},
  \bibnamefont{et~al.}, \bibinfo{journal}{Science}
  \textbf{\bibinfo{volume}{340}}, \bibinfo{pages}{167} (\bibinfo{year}{2013}).

\bibitem[{\citenamefont{Nandkishore et~al.}(2012)\citenamefont{Nandkishore,
  Levitov, and Chubukov}}]{Levitov_NatPhys2012}
\bibinfo{author}{\bibfnamefont{R.}~\bibnamefont{Nandkishore}},
  \bibinfo{author}{\bibfnamefont{L.~S.} \bibnamefont{Levitov}},
  \bibnamefont{and} \bibinfo{author}{\bibfnamefont{A.~V.}
  \bibnamefont{Chubukov}}, \bibinfo{journal}{Nat. Phys.}
  \textbf{\bibinfo{volume}{8}}, \bibinfo{pages}{158} (\bibinfo{year}{2012}).

\bibitem[{\citenamefont{Rosenbaum et~al.}(1983)\citenamefont{Rosenbaum,
  Milligan, Paalanen, Thomas, Bhatt, and Lin}}]{Rosenbaum_PRB1983}
\bibinfo{author}{\bibfnamefont{T.~F.} \bibnamefont{Rosenbaum}},
  \bibinfo{author}{\bibfnamefont{R.~F.} \bibnamefont{Milligan}},
  \bibinfo{author}{\bibfnamefont{M.~A.} \bibnamefont{Paalanen}},
  \bibinfo{author}{\bibfnamefont{G.~A.} \bibnamefont{Thomas}},
  \bibinfo{author}{\bibfnamefont{R.~N.} \bibnamefont{Bhatt}}, \bibnamefont{and}
  \bibinfo{author}{\bibfnamefont{W.}~\bibnamefont{Lin}},
  \bibinfo{journal}{Phys. Rev. B} \textbf{\bibinfo{volume}{27}},
  \bibinfo{pages}{7509} (\bibinfo{year}{1983}).

\bibitem[{\citenamefont{Dai et~al.}(1992)\citenamefont{Dai, Zhang, and
  Sarachik}}]{Sarachik_PRB1992}
\bibinfo{author}{\bibfnamefont{P.}~\bibnamefont{Dai}},
  \bibinfo{author}{\bibfnamefont{Y.}~\bibnamefont{Zhang}}, \bibnamefont{and}
  \bibinfo{author}{\bibfnamefont{M.~P.} \bibnamefont{Sarachik}},
  \bibinfo{journal}{Phys. Rev. B} \textbf{\bibinfo{volume}{45}},
  \bibinfo{pages}{3984} (\bibinfo{year}{1992}).

\bibitem[{\citenamefont{Lakner and
  L\"ohneysen}(1989{\natexlab{a}})}]{Loh_PRL1989}
\bibinfo{author}{\bibfnamefont{M.}~\bibnamefont{Lakner}} \bibnamefont{and}
  \bibinfo{author}{\bibfnamefont{H.~v.} \bibnamefont{L\"ohneysen}},
  \bibinfo{journal}{Phys. Rev. Lett.} \textbf{\bibinfo{volume}{63}},
  \bibinfo{pages}{648} (\bibinfo{year}{1989}{\natexlab{a}}).

\bibitem[{\citenamefont{Paalanen et~al.}(1988)\citenamefont{Paalanen, Graebner,
  Bhatt, and Sachdev}}]{Sachdev_Paalanen_PRL1988}
\bibinfo{author}{\bibfnamefont{M.~A.} \bibnamefont{Paalanen}},
  \bibinfo{author}{\bibfnamefont{J.~E.} \bibnamefont{Graebner}},
  \bibinfo{author}{\bibfnamefont{R.~N.} \bibnamefont{Bhatt}}, \bibnamefont{and}
  \bibinfo{author}{\bibfnamefont{S.}~\bibnamefont{Sachdev}},
  \bibinfo{journal}{Phys. Rev. Lett.} \textbf{\bibinfo{volume}{61}},
  \bibinfo{pages}{597} (\bibinfo{year}{1988}).

\bibitem[{\citenamefont{Sachdev}(1989)}]{Sachdev_Localmoment_PRB1989}
\bibinfo{author}{\bibfnamefont{S.}~\bibnamefont{Sachdev}},
  \bibinfo{journal}{Phys. Rev. B} \textbf{\bibinfo{volume}{39}},
  \bibinfo{pages}{5297} (\bibinfo{year}{1989}).

\bibitem[{\citenamefont{Milovanovi\ifmmode~\acute{c}\else \'{c}\fi{}
  et~al.}(1989)\citenamefont{Milovanovi\ifmmode~\acute{c}\else \'{c}\fi{},
  Sachdev, and Bhatt}}]{Sachdev_Bhatt_LocalMoment_PRL1989}
\bibinfo{author}{\bibfnamefont{M.}~\bibnamefont{Milovanovi\ifmmode~\acute{c}\e%
lse \'{c}\fi{}}}, \bibinfo{author}{\bibfnamefont{S.}~\bibnamefont{Sachdev}},
  \bibnamefont{and} \bibinfo{author}{\bibfnamefont{R.~N.} \bibnamefont{Bhatt}},
  \bibinfo{journal}{Phys. Rev. Lett.} \textbf{\bibinfo{volume}{63}},
  \bibinfo{pages}{82} (\bibinfo{year}{1989}).

\bibitem[{\citenamefont{Goh et~al.}(2006)\citenamefont{Goh, Oberbeck, Simmons,
  Hamilton, and Butcher}}]{Goh2006}
\bibinfo{author}{\bibfnamefont{K.~E.~J.} \bibnamefont{Goh}},
  \bibinfo{author}{\bibfnamefont{L.}~\bibnamefont{Oberbeck}},
  \bibinfo{author}{\bibfnamefont{M.~Y.} \bibnamefont{Simmons}},
  \bibinfo{author}{\bibfnamefont{A.~R.} \bibnamefont{Hamilton}},
  \bibnamefont{and} \bibinfo{author}{\bibfnamefont{M.~J.}
  \bibnamefont{Butcher}}, \bibinfo{journal}{Phys. Rev. B}
  \textbf{\bibinfo{volume}{73}}, \bibinfo{pages}{035401}
  (\bibinfo{year}{2006}).

\bibitem[{\citenamefont{Scappucci et~al.}(2012)\citenamefont{Scappucci, Klesse,
  Hamilton, Capellini, Jaeger, Bischof, Reidy, Gorman, and
  Simmons}}]{Giordano_Nanolett2012}
\bibinfo{author}{\bibfnamefont{G.}~\bibnamefont{Scappucci}},
  \bibinfo{author}{\bibfnamefont{W.~M.} \bibnamefont{Klesse}},
  \bibinfo{author}{\bibfnamefont{A.~R.} \bibnamefont{Hamilton}},
  \bibinfo{author}{\bibfnamefont{G.}~\bibnamefont{Capellini}},
  \bibinfo{author}{\bibfnamefont{D.~L.} \bibnamefont{Jaeger}},
  \bibinfo{author}{\bibfnamefont{M.~R.} \bibnamefont{Bischof}},
  \bibinfo{author}{\bibfnamefont{R.~F.} \bibnamefont{Reidy}},
  \bibinfo{author}{\bibfnamefont{B.~P.} \bibnamefont{Gorman}},
  \bibnamefont{and} \bibinfo{author}{\bibfnamefont{M.~Y.}
  \bibnamefont{Simmons}}, \bibinfo{journal}{Nano Letters}
  \textbf{\bibinfo{volume}{12}}, \bibinfo{pages}{4953} (\bibinfo{year}{2012}).

\bibitem[{\citenamefont{Weber et~al.}(2012)\citenamefont{Weber, Mahapatra, Ryu,
  Lee, Fuhrer, Reusch, Thompson, Lee, Klimeck, Hollenberg
  et~al.}}]{BentScience}
\bibinfo{author}{\bibfnamefont{B.}~\bibnamefont{Weber}},
  \bibinfo{author}{\bibfnamefont{S.}~\bibnamefont{Mahapatra}},
  \bibinfo{author}{\bibfnamefont{H.}~\bibnamefont{Ryu}},
  \bibinfo{author}{\bibfnamefont{S.}~\bibnamefont{Lee}},
  \bibinfo{author}{\bibfnamefont{A.}~\bibnamefont{Fuhrer}},
  \bibinfo{author}{\bibfnamefont{T.~C.~G.} \bibnamefont{Reusch}},
  \bibinfo{author}{\bibfnamefont{D.~L.} \bibnamefont{Thompson}},
  \bibinfo{author}{\bibfnamefont{W.~C.~T.} \bibnamefont{Lee}},
  \bibinfo{author}{\bibfnamefont{G.}~\bibnamefont{Klimeck}},
  \bibinfo{author}{\bibfnamefont{L.~C.~L.} \bibnamefont{Hollenberg}},
  \bibnamefont{et~al.}, \bibinfo{journal}{Science}
  \textbf{\bibinfo{volume}{335}}, \bibinfo{pages}{64} (\bibinfo{year}{2012}).

\bibitem[{\citenamefont{Fuechsle et~al.}(2012)\citenamefont{Fuechsle, Miwa,
  Mahapatra, Ryu, Lee, Warschkow, Hollenberg, Klimeck, and
  Simmons}}]{Fuechsle2012}
\bibinfo{author}{\bibfnamefont{M.}~\bibnamefont{Fuechsle}},
  \bibinfo{author}{\bibfnamefont{J.~A.} \bibnamefont{Miwa}},
  \bibinfo{author}{\bibfnamefont{S.}~\bibnamefont{Mahapatra}},
  \bibinfo{author}{\bibfnamefont{H.}~\bibnamefont{Ryu}},
  \bibinfo{author}{\bibfnamefont{S.}~\bibnamefont{Lee}},
  \bibinfo{author}{\bibfnamefont{O.}~\bibnamefont{Warschkow}},
  \bibinfo{author}{\bibfnamefont{L.~C.~L.} \bibnamefont{Hollenberg}},
  \bibinfo{author}{\bibfnamefont{G.}~\bibnamefont{Klimeck}}, \bibnamefont{and}
  \bibinfo{author}{\bibfnamefont{M.~Y.} \bibnamefont{Simmons}},
  \bibinfo{journal}{Nature Nanotechnology} \textbf{\bibinfo{volume}{7}},
  \bibinfo{pages}{242} (\bibinfo{year}{2012}).

\bibitem[{\citenamefont{Byczuk et~al.}(2005)\citenamefont{Byczuk, Hofstetter,
  and Vollhardt}}]{Vollhardt_PRL2005}
\bibinfo{author}{\bibfnamefont{K.}~\bibnamefont{Byczuk}},
  \bibinfo{author}{\bibfnamefont{W.}~\bibnamefont{Hofstetter}},
  \bibnamefont{and}
  \bibinfo{author}{\bibfnamefont{D.}~\bibnamefont{Vollhardt}},
  \bibinfo{journal}{Phys. Rev. Lett.} \textbf{\bibinfo{volume}{94}},
  \bibinfo{pages}{056404} (\bibinfo{year}{2005}).

\bibitem[{\citenamefont{Denteneer et~al.}(1999)\citenamefont{Denteneer,
  Scalettar, and Trivedi}}]{NandiniTrivedi_PRL1999}
\bibinfo{author}{\bibfnamefont{P.~J.~H.} \bibnamefont{Denteneer}},
  \bibinfo{author}{\bibfnamefont{R.~T.} \bibnamefont{Scalettar}},
  \bibnamefont{and} \bibinfo{author}{\bibfnamefont{N.}~\bibnamefont{Trivedi}},
  \bibinfo{journal}{Phys. Rev. Lett.} \textbf{\bibinfo{volume}{83}},
  \bibinfo{pages}{4610} (\bibinfo{year}{1999}).

\bibitem[{\citenamefont{Kohno}(2012)}]{Kohno_PRL2012}
\bibinfo{author}{\bibfnamefont{M.}~\bibnamefont{Kohno}},
  \bibinfo{journal}{Phys. Rev. Lett.} \textbf{\bibinfo{volume}{108}},
  \bibinfo{pages}{076401} (\bibinfo{year}{2012}).

\bibitem[{\citenamefont{Nielsen and Bhatt}(2007)}]{BhatPRB2007}
\bibinfo{author}{\bibfnamefont{E.}~\bibnamefont{Nielsen}} \bibnamefont{and}
  \bibinfo{author}{\bibfnamefont{R.~N.} \bibnamefont{Bhatt}},
  \bibinfo{journal}{Phys. Rev. B} \textbf{\bibinfo{volume}{76}},
  \bibinfo{pages}{161202} (\bibinfo{year}{2007}).

\bibitem[{\citenamefont{Goh}(2006)}]{Johnson_PhDThesis}
\bibinfo{author}{\bibfnamefont{K.~E.~J.} \bibnamefont{Goh}},
  \bibinfo{journal}{PhD Thesis, University of New South Wales}
  (\bibinfo{year}{2006}).

\bibitem[{\citenamefont{Hwang and Das~Sarma}(2013)}]{SDSarma_SiP_PRB2013}
\bibinfo{author}{\bibfnamefont{E.~H.} \bibnamefont{Hwang}} \bibnamefont{and}
  \bibinfo{author}{\bibfnamefont{S.}~\bibnamefont{Das~Sarma}},
  \bibinfo{journal}{Phys. Rev. B} \textbf{\bibinfo{volume}{87}},
  \bibinfo{pages}{125411} (\bibinfo{year}{2013}).

\bibitem[{\citenamefont{Shklovskii and Efros}(Springer-{V}erlag, {B}erlin
  {H}eidelberg {N}ew {Y}ork {T}okyo, 1984)}]{Efros_Shklovskii_Book}
\bibinfo{author}{\bibfnamefont{B.~I.} \bibnamefont{Shklovskii}}
  \bibnamefont{and} \bibinfo{author}{\bibfnamefont{A.~L.} \bibnamefont{Efros}},
  \bibinfo{journal}{Electronic properties of doped semiconductors}
  (\bibinfo{year}{Springer-{V}erlag, {B}erlin {H}eidelberg {N}ew {Y}ork
  {T}okyo, 1984}).

\bibitem[{\citenamefont{Gornyi and Mirlin}(2003)}]{Interaction_MR_PRL}
\bibinfo{author}{\bibfnamefont{I.~V.} \bibnamefont{Gornyi}} \bibnamefont{and}
  \bibinfo{author}{\bibfnamefont{A.~D.} \bibnamefont{Mirlin}},
  \bibinfo{journal}{Phys. Rev. Lett.} \textbf{\bibinfo{volume}{90}},
  \bibinfo{pages}{076801} (\bibinfo{year}{2003}).

\bibitem[{\citenamefont{Ghosh and Raychaudhuri}(2000)}]{AGPRL2000}
\bibinfo{author}{\bibfnamefont{A.}~\bibnamefont{Ghosh}} \bibnamefont{and}
  \bibinfo{author}{\bibfnamefont{A.~K.} \bibnamefont{Raychaudhuri}},
  \bibinfo{journal}{Phys. Rev. Lett.} \textbf{\bibinfo{volume}{84}},
  \bibinfo{pages}{4681} (\bibinfo{year}{2000}).

\bibitem[{\citenamefont{Dai et~al.}(1993)\citenamefont{Dai, Zhang, Bogdanovich,
  and Sarachik}}]{Critical_exponent_SiP_PRB}
\bibinfo{author}{\bibfnamefont{P.}~\bibnamefont{Dai}},
  \bibinfo{author}{\bibfnamefont{Y.}~\bibnamefont{Zhang}},
  \bibinfo{author}{\bibfnamefont{S.}~\bibnamefont{Bogdanovich}},
  \bibnamefont{and} \bibinfo{author}{\bibfnamefont{M.~P.}
  \bibnamefont{Sarachik}}, \bibinfo{journal}{Phys. Rev. B}
  \textbf{\bibinfo{volume}{48}}, \bibinfo{pages}{4941} (\bibinfo{year}{1993}).

\bibitem[{\citenamefont{Rahimi et~al.}(2003)\citenamefont{Rahimi, Anissimova,
  Sakr, Kravchenko, and Klapwijk}}]{Kravcehnko_PRL2003}
\bibinfo{author}{\bibfnamefont{M.}~\bibnamefont{Rahimi}},
  \bibinfo{author}{\bibfnamefont{S.}~\bibnamefont{Anissimova}},
  \bibinfo{author}{\bibfnamefont{M.~R.} \bibnamefont{Sakr}},
  \bibinfo{author}{\bibfnamefont{S.~V.} \bibnamefont{Kravchenko}},
  \bibnamefont{and} \bibinfo{author}{\bibfnamefont{T.~M.}
  \bibnamefont{Klapwijk}}, \bibinfo{journal}{Phys. Rev. Lett.}
  \textbf{\bibinfo{volume}{91}}, \bibinfo{pages}{116402}
  (\bibinfo{year}{2003}).

\bibitem[{\citenamefont{Imada et~al.}(1998)\citenamefont{Imada, Fujimori, and
  Tokura}}]{Imada_RMP1998}
\bibinfo{author}{\bibfnamefont{M.}~\bibnamefont{Imada}},
  \bibinfo{author}{\bibfnamefont{A.}~\bibnamefont{Fujimori}}, \bibnamefont{and}
  \bibinfo{author}{\bibfnamefont{Y.}~\bibnamefont{Tokura}},
  \bibinfo{journal}{Rev. Mod. Phys.} \textbf{\bibinfo{volume}{70}},
  \bibinfo{pages}{1039} (\bibinfo{year}{1998}).

\bibitem[{\citenamefont{Stone}(1989)}]{Stone_PRB}
\bibinfo{author}{\bibfnamefont{A.~D.} \bibnamefont{Stone}},
  \bibinfo{journal}{Phys. Rev. B} \textbf{\bibinfo{volume}{39}},
  \bibinfo{pages}{10736} (\bibinfo{year}{1989}).

\bibitem[{\citenamefont{Birge et~al.}(1990)\citenamefont{Birge, Golding, and
  Haemmerle}}]{BirgePRB1990}
\bibinfo{author}{\bibfnamefont{N.~O.} \bibnamefont{Birge}},
  \bibinfo{author}{\bibfnamefont{B.}~\bibnamefont{Golding}}, \bibnamefont{and}
  \bibinfo{author}{\bibfnamefont{W.~H.} \bibnamefont{Haemmerle}},
  \bibinfo{journal}{Phys. Rev. B} \textbf{\bibinfo{volume}{42}},
  \bibinfo{pages}{2735} (\bibinfo{year}{1990}).

\bibitem[{\citenamefont{McConville and Birge}(1993)}]{BirgePRB1993}
\bibinfo{author}{\bibfnamefont{P.}~\bibnamefont{McConville}} \bibnamefont{and}
  \bibinfo{author}{\bibfnamefont{N.~O.} \bibnamefont{Birge}},
  \bibinfo{journal}{Phys. Rev. B} \textbf{\bibinfo{volume}{47}},
  \bibinfo{pages}{16667} (\bibinfo{year}{1993}).

\bibitem[{\citenamefont{Feng et~al.}(1986)\citenamefont{Feng, Lee, and
  Stone}}]{FengPRL1986}
\bibinfo{author}{\bibfnamefont{S.}~\bibnamefont{Feng}},
  \bibinfo{author}{\bibfnamefont{P.~A.} \bibnamefont{Lee}}, \bibnamefont{and}
  \bibinfo{author}{\bibfnamefont{A.~D.} \bibnamefont{Stone}},
  \bibinfo{journal}{Phys. Rev. Lett.} \textbf{\bibinfo{volume}{56}},
  \bibinfo{pages}{1960} (\bibinfo{year}{1986}).

\bibitem[{\citenamefont{Shamim et~al.}(2011)\citenamefont{Shamim, Mahapatra,
  Polley, Simmons, and Ghosh}}]{SaquibPRB2011}
\bibinfo{author}{\bibfnamefont{S.}~\bibnamefont{Shamim}},
  \bibinfo{author}{\bibfnamefont{S.}~\bibnamefont{Mahapatra}},
  \bibinfo{author}{\bibfnamefont{C.}~\bibnamefont{Polley}},
  \bibinfo{author}{\bibfnamefont{M.~Y.} \bibnamefont{Simmons}},
  \bibnamefont{and} \bibinfo{author}{\bibfnamefont{A.}~\bibnamefont{Ghosh}},
  \bibinfo{journal}{Phys. Rev. B} \textbf{\bibinfo{volume}{83}},
  \bibinfo{pages}{233304} (\bibinfo{year}{2011}).

\bibitem[{\citenamefont{Altshuler and Spivak}(1985)}]{Altshuler_Spivak_1985}
\bibinfo{author}{\bibfnamefont{B.~L.} \bibnamefont{Altshuler}}
  \bibnamefont{and} \bibinfo{author}{\bibfnamefont{B.~Z.}
  \bibnamefont{Spivak}}, \bibinfo{journal}{Pis'ma Zh. Eksp. Teor. Fiz.}
  \textbf{\bibinfo{volume}{42}}, \bibinfo{pages}{363} (\bibinfo{year}{1985}).

\bibitem[{\citenamefont{Beenakker and van
  Houten}(1991)}]{Beenakker_Houten_Review}
\bibinfo{author}{\bibfnamefont{C.}~\bibnamefont{Beenakker}} \bibnamefont{and}
  \bibinfo{author}{\bibfnamefont{H.}~\bibnamefont{van Houten}}, in
  \emph{\bibinfo{booktitle}{Semiconductor Heterostructures and Nanostructures}}
  (\bibinfo{publisher}{Academic Press}, \bibinfo{year}{1991}),
  vol.~\bibinfo{volume}{44} of \emph{\bibinfo{series}{Solid State Physics}},
  pp. \bibinfo{pages}{1 -- 228}.

\bibitem[{\citenamefont{Moon et~al.}(1996)\citenamefont{Moon, Birge, and
  Golding}}]{BirgePRB1996_Zeeman}
\bibinfo{author}{\bibfnamefont{J.~S.} \bibnamefont{Moon}},
  \bibinfo{author}{\bibfnamefont{N.~O.} \bibnamefont{Birge}}, \bibnamefont{and}
  \bibinfo{author}{\bibfnamefont{B.}~\bibnamefont{Golding}},
  \bibinfo{journal}{Phys. Rev. B} \textbf{\bibinfo{volume}{53}},
  \bibinfo{pages}{R4193} (\bibinfo{year}{1996}).

\bibitem[{\citenamefont{Clarke et~al.}(2008)\citenamefont{Clarke, Yasin,
  Hamilton, Micolich, Simmons, Muraki, Hirayama, Pepper, and
  Ritchie}}]{MYS_NatPhys2008}
\bibinfo{author}{\bibfnamefont{W.~R.} \bibnamefont{Clarke}},
  \bibinfo{author}{\bibfnamefont{C.~E.} \bibnamefont{Yasin}},
  \bibinfo{author}{\bibfnamefont{A.~R.} \bibnamefont{Hamilton}},
  \bibinfo{author}{\bibfnamefont{A.~P.} \bibnamefont{Micolich}},
  \bibinfo{author}{\bibfnamefont{M.~Y.} \bibnamefont{Simmons}},
  \bibinfo{author}{\bibfnamefont{K.}~\bibnamefont{Muraki}},
  \bibinfo{author}{\bibfnamefont{Y.}~\bibnamefont{Hirayama}},
  \bibinfo{author}{\bibfnamefont{M.}~\bibnamefont{Pepper}}, \bibnamefont{and}
  \bibinfo{author}{\bibfnamefont{D.~A.} \bibnamefont{Ritchie}},
  \bibinfo{journal}{Nat. Phys.} \textbf{\bibinfo{volume}{4}},
  \bibinfo{pages}{55} (\bibinfo{year}{2008}).

\bibitem[{\citenamefont{Abrahams et~al.}(2001)\citenamefont{Abrahams,
  Kravchenko, and Sarachik}}]{Kravchenko_RMP2001}
\bibinfo{author}{\bibfnamefont{E.}~\bibnamefont{Abrahams}},
  \bibinfo{author}{\bibfnamefont{S.~V.} \bibnamefont{Kravchenko}},
  \bibnamefont{and} \bibinfo{author}{\bibfnamefont{M.~P.}
  \bibnamefont{Sarachik}}, \bibinfo{journal}{Rev. Mod. Phys.}
  \textbf{\bibinfo{volume}{73}}, \bibinfo{pages}{251} (\bibinfo{year}{2001}).

\bibitem[{\citenamefont{{Meyer} et~al.}(2002)\citenamefont{{Meyer}, {Fal'ko},
  and {Altshuler}}}]{InplaneMR_Meyer_arxiv}
\bibinfo{author}{\bibfnamefont{J.~S.} \bibnamefont{{Meyer}}},
  \bibinfo{author}{\bibfnamefont{V.~I.} \bibnamefont{{Fal'ko}}},
  \bibnamefont{and} \bibinfo{author}{\bibfnamefont{B.~L.}
  \bibnamefont{{Altshuler}}}, \bibinfo{journal}{eprint arXiv:cond-mat/0206024}
  (\bibinfo{year}{2002}), \eprint{arXiv:cond-mat/0206024}.

\bibitem[{\citenamefont{Christianson et~al.}(2002)\citenamefont{Christianson,
  Lacerda, Hundley, Pagliuso, and Sarrao}}]{MR_CeRhIn_PRB2002}
\bibinfo{author}{\bibfnamefont{A.~D.} \bibnamefont{Christianson}},
  \bibinfo{author}{\bibfnamefont{A.~H.} \bibnamefont{Lacerda}},
  \bibinfo{author}{\bibfnamefont{M.~F.} \bibnamefont{Hundley}},
  \bibinfo{author}{\bibfnamefont{P.~G.} \bibnamefont{Pagliuso}},
  \bibnamefont{and} \bibinfo{author}{\bibfnamefont{J.~L.}
  \bibnamefont{Sarrao}}, \bibinfo{journal}{Phys. Rev. B}
  \textbf{\bibinfo{volume}{66}}, \bibinfo{pages}{054410}
  (\bibinfo{year}{2002}).

\bibitem[{\citenamefont{Lakner and
  L\"ohneysen}(1989{\natexlab{b}})}]{Lohneysen_SpecificHeat_PRL1989}
\bibinfo{author}{\bibfnamefont{M.}~\bibnamefont{Lakner}} \bibnamefont{and}
  \bibinfo{author}{\bibfnamefont{H.~v.} \bibnamefont{L\"ohneysen}},
  \bibinfo{journal}{Phys. Rev. Lett.} \textbf{\bibinfo{volume}{63}},
  \bibinfo{pages}{648} (\bibinfo{year}{1989}{\natexlab{b}}).

\bibitem[{\citenamefont{Punnoose and
  Finkel'stein}(2005)}]{Punnoose_Science2005}
\bibinfo{author}{\bibfnamefont{A.}~\bibnamefont{Punnoose}} \bibnamefont{and}
  \bibinfo{author}{\bibfnamefont{A.~M.} \bibnamefont{Finkel'stein}},
  \bibinfo{journal}{Science} \textbf{\bibinfo{volume}{310}},
  \bibinfo{pages}{289} (\bibinfo{year}{2005}).

\bibitem[{\citenamefont{Gunawan et~al.}(2006)\citenamefont{Gunawan, Gokmen,
  Vakili, Padmanabhan, De~Poortere, and Shayegan}}]{Medini_NatPhys2006}
\bibinfo{author}{\bibfnamefont{O.}~\bibnamefont{Gunawan}},
  \bibinfo{author}{\bibfnamefont{T.}~\bibnamefont{Gokmen}},
  \bibinfo{author}{\bibfnamefont{K.}~\bibnamefont{Vakili}},
  \bibinfo{author}{\bibfnamefont{M.}~\bibnamefont{Padmanabhan}},
  \bibinfo{author}{\bibfnamefont{E.~P.} \bibnamefont{De~Poortere}},
  \bibnamefont{and} \bibinfo{author}{\bibfnamefont{M.}~\bibnamefont{Shayegan}},
  \bibinfo{journal}{Nat. Phys.} \textbf{\bibinfo{volume}{3}},
  \bibinfo{pages}{388} (\bibinfo{year}{2006}).

\bibitem[{\citenamefont{Shepelyansky}(1994)}]{Shepelyansky_PRL1994}
\bibinfo{author}{\bibfnamefont{D.~L.} \bibnamefont{Shepelyansky}},
  \bibinfo{journal}{Phys. Rev. Lett.} \textbf{\bibinfo{volume}{73}},
  \bibinfo{pages}{2607} (\bibinfo{year}{1994}).

\end{thebibliography}
\end{document}